\documentclass[aps,prb,twocolumn,superscriptaddress,longbibliography]{revtex4-2}

\usepackage{amssymb,amsmath,amsthm,bbm,bbold}
\usepackage{amsfonts}
\usepackage{graphicx}
\usepackage{mathtools}
\usepackage[breaklinks,colorlinks,allcolors=blue]{hyperref}
\usepackage[normalem]{ulem}
\usepackage[T1]{fontenc}
\usepackage{color}
\usepackage{xcolor}
\usepackage{tcolorbox}

\usepackage{pifont}
\usepackage{amssymb}
\usepackage{tikz}
\usepackage{makecell}

\def\CC{\mathbbm{C}}

\newcommand{\bd}[1]{\boldsymbol{#1}}

\usepackage{accents}

\newcommand{\Tr}{\mathrm{Tr}}

\newcommand*\xbar[1]{\hbox{\vbox{
       \hrule height 0.6pt 
       \kern0.3ex
       \hbox{%
         \kern-0.2em
         \ensuremath{#1}%
         \kern 0.0em
         }}}}

\newcommand*\xxbar[1]{\hbox{\vbox{
       \hrule height 0.6pt 
       \kern0.3ex
       \hbox{%
         \kern-0.0em
         \ensuremath{#1}%
         \kern 0.0em
         }}}}

\newcommand{\bra}[1]{\mbox{$\langle #1 |$}}
\newcommand{\ket}[1]{\mbox{$| #1 \rangle$}}

\begin{document}
\title{Spin adaptation of the cumulant expansions of reduced density matrices}
\author{Julia Liebert}
\affiliation{Department of Physics, Arnold Sommerfeld Center for Theoretical Physics, Ludwig-Maximilians-Universität München, Theresienstrasse 37, 80333 München, Germany}
\affiliation{Munich Center for Quantum Science and Technology (MCQST), Schellingstrasse 4, 80799 München, Germany}
\author{Christian Schilling}
\affiliation{Department of Physics, Arnold Sommerfeld Center for Theoretical Physics, Ludwig-Maximilians-Universität München, Theresienstrasse 37, 80333 München, Germany}
\affiliation{Munich Center for Quantum Science and Technology (MCQST), Schellingstrasse 4, 80799 München, Germany}
\author{David A. Mazziotti}
\email{damazz@uchicago.edu}
\affiliation{Department of Chemistry and The James Franck Institute, The University of Chicago, Chicago, IL 60637}
\date{Submitted May 22, 2025}

\begin{abstract}
We develop a systematic framework for the spin adaptation of the cumulants of $p$-particle reduced density matrices (RDMs), with explicit constructions for $p = 1$ to $3$. These spin-adapted cumulants enable rigorous treatment of both $\hat{S}_z$ and $\hat{S}^2$ symmetries in quantum systems, providing a foundation for spin-resolved electronic structure methods. We show that \textit{complete} spin adaptation---referred to as \textit{complete $S$-representability}---can be enforced by constraining the variances of $\hat{S}_z$ and $\hat{S}^2$, which require the 2-RDM and 4-RDM, respectively. Importantly, the cumulants of RDMs scale linearly with system size---\textit{size-extensive}---making them a natural object for incorporating spin symmetries in scalable electronic structure theories. The developed formalism is applicable to density-based methods (DFT), one-particle RDM functional theories (RDMFT), and two-particle RDM methods. We further extend the approach to spin-orbit-coupled systems via total angular momentum adaptation. Beyond spin, the framework enables the adaptation of RDM theories to additional symmetries through the construction of suitable irreducible tensor operators.
\end{abstract}


\maketitle

\section{Introduction\label{sec:intro}}

Reduced density matrices (RDMs) play a central role in modern quantum chemistry \cite{D76, C00, M06-review}, offering a promising approach to bypass the exponential scaling of the many-particle Hilbert space with the system size. A key example is Hartree-Fock theory, which can be formulated in terms of the one-particle reduced density matrix (1RDM) \cite{Slater51, VM14, NDP18}.
At the one-particle level, Kohn-Sham density functional theory (KS-DFT) \cite{KS65, GD95} usually fails to describe systems with strong static correlation, a limitation that can be addressed by other RDM methods. In particular, one-particle reduced density matrix functional theory (RDMFT) has been shown to effectively capture strong correlation while remaining computationally less expensive than wave function methods \cite{PG16, SB18, Piris19, P21,DVR21, LCLS21, LKO22,GBM22,MP22,SNF22,LdCP23, GBM23, LCS23, LS23-sp, LS23-njp, CG24, YS24, GBM24, VMSS24}.
In two-particle reduced density matrix (2RDM) methods, the need to construct universal functionals as in DFT or RDMFT is eliminated, as the expectation value of any pairwise interacting Hamiltonian simplifies to a linear functional of the 2RDM \cite{M04, M05, CL06, VAVAB09, NBFFP08, LBSIB15, M20, DLBBKRB23, DP24}. The energy minimization is then performed under $N$-representability conditions, ensuring that the 2RDM corresponds to a valid $N$-fermion quantum state \cite{C63, Kummer67, LCV07, M12-1, M12-2, M23}.

Furthermore, if the total spin $S$ is a good quantum number, enforcing $S$-representability of 2RDMs in variational ground state calculations is crucial to avoid spin contamination \cite{aszabo89, Krylov17}. Similar to $N$-representability, $S$-representability of a $p$-particle reduced density matrix ($p$-RDM) ensures compatibility with an $N$-fermion state of well-defined $S$, with an analogous condition applying to the magnetization $M$. Since the Casimir operator $\boldsymbol{S}^2$ is a two-particle operator, a correct expectation value of $\boldsymbol{S}^2$ can be ensured on the two-particle level.
However, fixing the expectation value is only necessary but not sufficient to guarantee $S$-representability \cite{PRTV97, GM05} of a 2RDM.

In RDM methods, correct spin quantum numbers can be ensured either by a spin-free formalism \cite{K99,SKR09, KSD10} or through direct spin adaptation of the RDMs using irreducible tensor operators \cite{GM05, Mazziotti.2005, vAVBvNA12}. Unlike the spin-free formalism \cite{K99, SKR09, KSD10}, spin adaptation is not restricted to spin-independent Hamiltonians. This is crucial, as spin-dependent Hamiltonians are essential in cases such as spin-orbit coupling \cite{McWeeny65, GB93-so, GB93, BG96, YHAE95, Marian01} or systems with external or internal magnetic fields \cite{NMR04, VC05, BDZ08, FBO12,vK86, vKDP80, ES90, Huckestein95, STG99, MR03, BZ06, CLMD23}. For spin-dependent Hamiltonians, additional blocks of the spin-adapted $p$-RDMs may be nonzero, allowing direct investigation of their contribution to the ground-state energy, which the spin-free formalism cannot provide.

In this paper, we develop a systematic framework for the spin adaptation of the cumulants of $p$-RDMs. This spin-adapted formalism resolves a key limitation of its spin-free counterpart~\cite{KSD10} by providing a clear and tractable approach to extending the theory to higher-order $p$-particle cumulants. Specifically, we show how spin-adapted $p$-particle cumulants naturally emerge from spin-adapted RDMs. We derive explicit conditions for their traces in terms of the spin quantum numbers $S,M$ for spin-adapted cumulants up to the four-particle level. The full $S$-representability of RDMs and their cumulants can be enforced by requiring that the variance of the Casimir operator $\boldsymbol{S}^2$, expressible in terms of the 4-RDM, vanishes.  Finally, we demonstrate that this spin-adapted framework generalizes to the total angular momentum representation relevant for systems with spin-orbit coupling.

The paper is structured as follows: In Sec.~\ref{sec:one} we discuss spin adaptation and conservation of magnetization on the one-particle level, where the 1RDM and its cumulant coincide. In Sec.~\ref{sec:two}, we introduce spin-adapted cumulants of the 2RDM and explore their role in enforcing the correct expectation value of $\boldsymbol{S}^2$. We eventually extend our results to a generic $p$-particle level and focus on the explicit constraints on the spin-adapted cumulants on the three-particle level in Sec.~\ref{sec:three}. Finally, in Sec.~\ref{sec:four}, we discuss spin adaptation at the four-particle level before we discuss its extension to the total angular momentum representation in Sec.~\ref{sec:LS}.

\section{Theory \label{sec:theory}}

In the following, we present a comprehensive framework for incorporating spin symmetries into RDM theories, with a focus on the spin adaptation of $p$-particle RDM cumulants and the $S$-representability problem.

\subsection{Conservation of magnetization \label{sec:one}}

In this section, we briefly review spin adaptation at the one-particle level and introduce the notion of irreducible tensor operators, with a focus on the conservation of total magnetization, as it is a one-particle symmetry.
The one-particle Hilbert space of spin-$1/2$ fermions with spin $\sigma\in\{\alpha, \beta\}$ reads
\begin{equation}
\mathcal{H}_1 = \mathcal{H}_1^{(l)}\oplus \mathcal{H}_1^{(l)}\cong \CC^d\otimes \CC^2\,,
\end{equation}
where $\mathcal{H}_1^{(l)}$ denotes Hilbert space associated with the (spatial) orbital degrees of freedom and $d=\mathrm{dim}(\mathcal{H}_1^{(l)})<\infty$.

Conservation of total magnetization corresponds to a global $\text{U}(1)$ symmetry. The infinitesimal generator of the symmetry group is the $S_z$ operator,
\begin{equation}\label{eq:Sz-operator}
S_z =\frac{1}{2} \sum_{i=1}^d f_{i\alpha}^\dagger f_{i\alpha}-f_{i\beta}^\dagger f_{i\beta}\,,
\end{equation}
where $f_{i\sigma}^\dagger$, $(f_{i\sigma})$ denote the fermionic creation (annihilation) operators of a particle with spin $\sigma \in\{\alpha, \beta\}$. For conserved $M = \langle S_z\rangle$, the 1RDM ${}^1D= {}^1D^{\alpha}_\alpha\oplus {}^1D^\beta_\beta$ is block-diagonal with $({}^1D^\sigma_\sigma)_{ij} = \mathrm{Tr}[f_{j\sigma}^\dagger f_{i\sigma}], \sigma\in\{\alpha, \beta\}$.

The spin-adapted one-particle creation operators ${}^1C^{\lambda, \mu}$ given by ${}^1C^{1/2, 1/2}_i = f_{i\alpha}^\dagger \,,{}^1C^{1/2, -1/2}_i = f_{i\beta}^\dagger$ \cite{Harriman79} provide a basis for the 1RDM.
Here, the superscript $\lambda$ refers to the total spin and $\mu$ to the magnetization. We use $\lambda, \mu$ instead of $S,M$ to avoid confusion with the system's total spin $S$ and magnetization $M$.
Furthermore, the operators ${}^1C^{\lambda,\mu}_i$ are the components of $p$-particle irreducible spin tensor operators of rank $\lambda$ for the case $p=1$, satisfying the defining properties \cite{Harriman79, McWeeny61, Edmonds1957}:
\begin{align}\label{eq:tensor-op}
[S_{\pm}, {}^pC^{\lambda,\mu}] &= \sqrt{\lambda(\lambda+1)-\mu(\mu\pm 1)}\, \,{}^pC^{\lambda,\mu\pm 1}\,,\nonumber\\
[S_z,{}^pC^{\lambda,\mu}]  &= \mu\,\, {}^pC^{\lambda,\mu}\,.
\end{align}
The resulting spin-adapted basis $\{{}^1(C^{\lambda, \mu}_i)^\dagger\ket{\Psi}|\}_{i=1}^d$ leads to the spin-adapted 1RDM matrix elements
\begin{equation}\label{eq:1RDM-adapt}
\left({}^1D^{\lambda \mu}_{\lambda^\prime \mu^\prime}\right)^j_i \equiv \bra{\Psi}{}^1C^{\lambda, \mu}_j\left({}^1C^{\lambda^\prime, \mu^\prime}_i\right)^\dagger\ket{\Psi}
\end{equation}
for $\lambda=\lambda^\prime=1/2$ and $\mu, \mu^\prime = \pm 1/2$.
Since the 1RDM cumulant ${}^1\Delta$ and the 1RDM ${}^1D$ coincide \cite{M99, K99}, the spin-adapted cumulant matrix elements directly follow from Eq.~\eqref{eq:1RDM-adapt} as
\begin{equation}
\left({}^1\Delta^{\lambda \mu}_{\lambda^\prime \mu^\prime}\right)^j_i = \left({}^1D^{\lambda \mu}_{\lambda^\prime \mu^\prime}\right)^j_i\,.
\end{equation}

On the level of the spin-adapted 1RDM, conservation of $M$ implies the following trace condition on the spin-adapted matrix elements,
\begin{equation}\label{eq:condition-M}
2M = \Tr\left[{}^1D^{1/2\,1/2}_{1/2\,1/2}\right] -\Tr\left[{}^1D^{1/2 \,-1/2}_{1/2\, -1/2}\right]\,.
\end{equation}
Thus, the correct expectation value of $S_z$ can be conveniently imposed on the one-particle level. In particular, this resembles the common observation that $M$ can be easily taken into account in any RDM functional theory, including DFT and RDMFT.
However, to ensure that an $N$-fermion state $\ket{\Psi}$ has a well-defined total magnetization $M$, Eq.~\eqref{eq:condition-M} alone is insufficient. For $M$-representability, the variance of the $S_z$ operator \eqref{eq:Sz-operator} must also vanish.

\subsection{Inclusion of total $S$ and spin-adapted 2RDM cumulants \label{sec:two}}

The advantage of having access to the 2RDM in the case of spin symmetries is two-fold. First, vanishing of the variance $\Delta S_z = \langle S_z^2\rangle - \langle S_z\rangle^2$ ensures that a pure state $N$-representable 2RDM corresponds to a quantum state with a well-defined $M$ quantum number. Second, it allows us to control the expectation value of the total spin $S$ via linear trace conditions on the spin-adapted 2RDM blocks. The total spin $S$ of a quantum state follows from the expectation value $\langle\bd S^2\rangle = S(S+1)$ of the Casimir operator
\begin{equation}\label{eq:S2}
\bd S^2 = S_z^2 +\frac{1}{2}\left(S_+ S_- +S_- S_+\right)\,,
\end{equation}
where $S_+ = \sum_{i=1}^df_{i\alpha}^\dagger f_{i\beta}$ ($S_- = \sum_{i=1}^d f_{i\beta}^\dagger f_{i\alpha}$) are the spin raising (lowering) operators. In the following, we first spin adapt the 2RDM in order to express the expectation value of $\bd S^2$ in terms of the spin-adapted 2RDM matrix elements.
The spin-adapted two-particle creation operators are given by the components of the irreducible tensor operators \cite{PRTV97, GM05}
\begin{align}\label{eq:tos-2rdm}
{}^2C^{0,0}_{ij}&= \frac{1}{\sqrt{2}}\left(f_{j\alpha}^\dagger f_{i\beta}^\dagger - f_{j\beta}^\dagger f_{i\alpha}^\dagger\right)\,,\nonumber\\
{}^2C^{1,0}_{ij} &= \frac{1}{\sqrt{2}}\left(f_{j\alpha}^\dagger f_{i\beta}^\dagger + f_{j\beta}^\dagger f_{i\alpha}^\dagger\right)\,,\nonumber\\
{}^2C^{1,1}_{ij} &= f_{j\alpha}^\dagger f_{i\alpha}^\dagger\,,\quad {}^2C^{1,-1}_{ij} = f_{j\beta}^\dagger f_{i\beta}^\dagger\,.
\end{align}
Then, the spin-adapted 2RDM matrix elements are defined in analogy to the spin-adapted 1RDM matrix elements in Eq.~\eqref{eq:1RDM-adapt} as \cite{GM05}
\begin{equation}\label{eq:2rdm-general}
\left( ^2D^{\lambda\mu}_{\lambda^\prime\mu^\prime}\right)^{ij}_{kl}\equiv \bra{\Psi}C^{\lambda\mu}_{ij} \left(C^{\lambda^\prime\mu^\prime}_{kl}\right)^\dagger\ket{\Psi}\,.
\end{equation}
The adjoint of the elements $C^{\lambda, \mu}$ of an irreducible tensor operators satisfy the relation $(C^{\lambda, \mu})^\dagger = (-1)^\mu C^{\lambda, -\mu}$.
Therefore, the spin-adapted 2RDM matrix elements are expressible in terms of expectation values of irreducible tensor operators as
\begin{equation}\label{eq:CG-product}
C^{\lambda_1, \mu_1} \left(C^{\lambda_2, \mu_2}\right)^\dagger = (-1)^{\mu_2} \sum_\lambda\begin{bmatrix}
\lambda_1 &\lambda_2&\lambda  \\ \mu_1& -\mu_2& \mu
\end{bmatrix}C^{\lambda, \mu}\,,
\end{equation}
where the sum runs over all $\lambda\in\{|\lambda_1-\lambda_2|, ..., \lambda_1+\lambda_2\}$ and the bracket denotes the Clebsch-Gordon coefficient \cite{Edmonds1957}.
Moreover, the Wigner-Eckart theorem states that for initial and final states $|\Psi_{S_i,M_i}\rangle \in\mathcal{H}_N^{(S_i, M_i)},|\Psi_{S_f,M_f}\rangle\in\mathcal{H}_N^{(S_f, M_f)}$ and an irreducible tensor operator $C^{\lambda,\mu}$ \cite{Hall15, Edmonds1957}:
\begin{equation}\label{eq:WE}
\langle \Psi_{S_f,M_f}|C^{\lambda, \mu}|\Psi_{S_i,M_i}\rangle = \begin{bmatrix}
S_i&\lambda &S_f\\ M_i& \mu & M_f
\end{bmatrix}\langle\Psi_{S_f}|| C^{\lambda,\mu}||\Psi_{S_i}\rangle\,,
\end{equation}
where $\langle\Psi_{S_f}|| C^{\lambda,\mu}||\Psi_{S_i}\rangle$ denotes the reduced matrix element that depends only on $S_{f}, S_i$. The Clebsch-Gordon coefficients in Eq.~\eqref{eq:WE} vanish unless $M_i+\mu=M_f$. Therefore, the Wigner-Eckart theorem directly implies that the 2RDM blocks that couple two operators ${}^2C^{\lambda, \mu}, ({}^2C^{\lambda^\prime, \mu^\prime})^\dagger$ with $\mu\neq \mu^\prime$ vanish for $M$ conserved. Hence, for $M$ conserved there are only six non-vanishing spin-adapted 2RDM blocks given by \cite{GM05}
\begin{align}\label{eq:2RDM-adapt}
\left( {}^2D^{00}_{00}\right)^{ij}_{kl} &= \frac{1}{2}\left({}^2D^{i\beta, j\alpha}_{k\beta, l\alpha} + {}^2D^{i\alpha, j\beta}_{k\alpha, l\beta}- {}^2D^{i\alpha, j\beta}_{k\beta, l\alpha}-{}^2D^{i\beta, j\alpha}_{k\alpha, l\beta}\right),\nonumber\\
\left( {}^2D^{10}_{10}\right)^{ij}_{kl} &= \frac{1}{2}\left({}^2 D^{i\beta, j\alpha}_{k\beta, l\alpha} + {}^2D^{i\alpha, j\beta}_{k\alpha, l\beta}+ {}^2D^{i\alpha, j\beta}_{k\beta, l\alpha} +{}^2D^{i\beta, j\alpha}_{k\alpha, l\beta}\right),\nonumber\\
\left( {}^2D^{00}_{10}\right)^{ij}_{kl} &= \frac{1}{2}\left({}^2D^{i\beta, j\alpha}_{k\beta, l\alpha} + {}^2D^{i\beta, j\alpha}_{k\alpha, l\beta}- {}^2D^{i\alpha, j\beta}_{k\beta, l\alpha}-{}^2D^{i\alpha, j\beta}_{k\alpha, l\beta}\right),\nonumber\\
\left( {}^2D^{10}_{00}\right)^{ij}_{kl} &= \frac{1}{2}\left({}^2D^{i\beta, j\alpha}_{k\beta, l\alpha} + {}^2D^{i\alpha, j\beta}_{k\beta, l\alpha}- {}^2D^{i\beta, j\alpha}_{k\alpha, l\beta}-{}^2D^{i\alpha, j\beta}_{k\alpha, l\beta}\right),\nonumber\\
\left( {}^2D^{11}_{11}\right)^{ij}_{kl} &=  {}^2D^{i\alpha, j\alpha}_{k\alpha, l\alpha},\nonumber\\
\left( {}^2D^{1-1}_{1-1}\right)^{ij}_{kl} &=  {}^2D^{i\beta, j\beta}_{k\beta, l\beta}.
\end{align}
The two spin-adapted 2RDM blocks ${}^2D^{00}_{00}, {}^2D_{10}^{00}$ are symmetric under the exchange of two spatial indices, while the four blocks ${}^2D^{10}_{10}, {}^2D_{00}^{10}, {}^2D^{11}_{11}, {}^2D^{1-1}_{1-1}$ are antisymmetric.
In the special case of $M=0$, also the 2RDM blocks ${}^2D^{00}_{10}, {}^2D^{10}_{00}$ with the same $\mu$ but different $\lambda$'s vanish.
Moreover, the expectation value of the Casimir operator $\bd S^2$ in Eq.~\eqref{eq:S2} is expressible in terms of the remaining four spin-adapted 2RDM blocks according to
\begin{equation}\label{eq:S2-expec}
\langle \bd S^2\rangle = \langle S_z^2\rangle+\frac{1}{2}\left(N+\Tr\left[{}^2D^{10}_{10}-{}^2D^{00}_{00}\right]\right)
\end{equation}
with
\begin{equation}\label{eq:Sz-expec}
\langle S_z^2\rangle = \frac{1}{4}\left(N+\Tr\left[{}^2D^{11}_{11}+{}^2D^{1-1}_{1-1}-{}^2D^{00}_{00}-{}^2D^{10}_{10}\right]\right)\,.
\end{equation}
In particular, the variance $\Delta S_z$ vanishes for $M$-representable 2RDMs and $\langle S_z^2\rangle = M^2$. From Eqs.~\eqref{eq:S2-expec} and \eqref{eq:Sz-expec} we further observe that the two blocks ${}^2D^{00}_{00}, {}^2D^{10}_{10}$ are constrained by the spin quantum numbers $S,M$, while ${}^2D^{11}_{11}, {}^2D^{1-1}_{1-1}$ only depend on the numbers $N_\alpha, N_\beta$ of particles with spin $\alpha, \beta$, respectively.
This leads to the following conditions on the traces of the spin-adapted 2RDM blocks (see also Refs.~\cite{PRTV97, GM05})
\begin{align}\label{eq:traces}
\Tr\left[{}^2D^{00}_{00}\right] &=\frac{N(N+2)}{4}-S(S+1)\,, \nonumber\\
\Tr\left[{}^2D^{10}_{10}\right] &= \frac{N(N-2)}{4} - 2M^2 +S(S+1)\,,  \nonumber\\
\Tr\left[{}^2D^{00}_{10}\right] &= \Tr_2\left[{}^2D^{10}_{00}\right]=0\,,\nonumber\\
\Tr\left[{}^2D^{11}_{11}\right] &= N_\alpha(N_\alpha-1)\,,\nonumber\\
\Tr\left[{}^2D^{1-1}_{1-1}\right] &= N_\beta(N_\beta-1)\,,
\end{align}
where the constraint on $\Tr[{}^2D^{00}_{10}]$ arises from $\Tr[{}^2D^{\alpha\beta}_{\beta\alpha}] = \Tr[{}^2D^{\beta\alpha}_{\alpha\beta}]$ and $\Tr[{}^2D^{\alpha\beta}_{\alpha\beta}] = \Tr[{}^2D^{\beta\alpha}_{\beta\alpha}]$.

Next, we eventually introduce spin-adapted cumulants for the 2RDM as
\begin{equation}\label{eq:cumulant-2rdm}
{}^2\Delta^{\lambda\mu}_{\lambda^\prime \mu^\prime}\equiv {}^2D^{\lambda \mu}_{\lambda^\prime \mu^\prime} - {}^2\kappa^{\lambda \mu}_{\lambda^\prime \mu^\prime}\,,
\end{equation}
where ${}^2\kappa^{\lambda \mu}_{\lambda^\prime \mu^\prime}$ is completely expressible in terms of the spin-adapted 1RDM matrix elements.
The explicit expressions for the spin-adapted non-cumulant exhibit the same structure as the spin-adapted 2RDMs themselves (recall Eq.~\eqref{eq:2RDM-adapt}) such that
\begin{align}\label{eq:cadapt-cum}
{}^2\Delta^{00}_{00} &= \frac{1}{2}\left({}^2\Delta^{\beta \alpha}_{\beta \alpha}+{}^2\Delta^{\alpha \beta}_{\alpha \beta}-{}^2\Delta^{\alpha \beta}_{\beta \alpha}-{}^2\Delta^{\beta \alpha}_{\alpha \beta}\right)\,,\nonumber\\
{}^2\Delta^{10}_{10} &=\frac{1}{2}\left({}^2\Delta^{\beta\alpha}_{\beta\alpha}+{}^2\Delta^{\alpha\beta}_{\alpha\beta}+{}^2\Delta^{\alpha\beta}_{\beta\alpha}+{}^2\Delta^{\beta\alpha}_{\alpha\beta}\right)\,,\nonumber\\
{}^2\Delta^{00}_{10} &=\frac{1}{2}\left({}^2\Delta^{\beta\alpha}_{\beta\alpha}-{}^2\Delta^{\alpha\beta}_{\alpha\beta}-{}^2\Delta^{\alpha\beta}_{\beta\alpha}+{}^2\Delta^{\beta\alpha}_{\alpha\beta}\right)\,,\nonumber\\
{}^2\Delta^{10}_{00} &=\frac{1}{2}\left({}^2\Delta^{\beta\alpha}_{\beta\alpha}-{}^2\Delta^{\alpha\beta}_{\alpha\beta}+{}^2\Delta^{\alpha\beta}_{\beta\alpha}-{}^2\Delta^{\beta\alpha}_{\alpha\beta}\right)\,,\nonumber\\
{}^2\Delta^{11}_{11} &={}^2\Delta^{\alpha\alpha}_{\alpha\alpha}\,,\nonumber\\
{}^2\Delta^{1-1}_{1-1} &={}^2\Delta^{\beta\beta}_{\beta\beta}\,.
\end{align}
Furthermore, the trace conditions for the spin-adapted 2RDM matrix elements \eqref{eq:traces} in combination with the non-cumulant parts ${}^2\kappa^{\lambda \mu}_{\lambda^\prime \mu^\prime}$ then imply trace conditions on the spin-adapted cumulants ${}^2\Delta^{\lambda\mu}_{\lambda^\prime\mu^\prime}$ in terms of $S, N, N_{\alpha}$ and $ N_{\beta}$ according to
\begin{equation}
\Tr\left[{}^2\Delta^{\lambda\mu}_{\lambda^\prime\mu^\prime}\right] = \Tr\left[ {}^2D^{\lambda\mu}_{\lambda^\prime\mu^\prime}\right]-\Tr\left[{}^2\kappa^{\lambda\mu}_{\lambda^\prime\mu^\prime}\right]\,.
\end{equation}
The required traces of the non-cumulant parts ${}^2\kappa^{\lambda \mu}_{\lambda^\prime \mu^\prime}$ of the cumulants in Eq.~\eqref{eq:cadapt-cum} are given by
\begin{align}\label{eq:traces-kappa}
\Tr\left[{}^2\kappa^{00}_{00}\right] &= N_\alpha N_\beta + \Tr\left[{}^1D^{1/2\,1/2}_{1/2\,1/2}\cdot {}^1D^{1/2\,-1/2}_{1/2\,-1/2} \right] \nonumber\\
&\quad - \Tr\left[{}^1D^{1/2\,1/2}_{1/2\,-1/2}\cdot {}^1D^{1/2\,-1/2}_{1/2\,1/2} \right] \nonumber\\
&\quad- \Tr\left[{}^1D^{1/2\,1/2}_{1/2\,-1/2}\right]\Tr\left[{}^1D^{1/2\,-1/2}_{1/2\,1/2}\right]  \,,\nonumber\\
\Tr\left[{}^2\kappa^{10}_{10}\right] &=  N_\alpha N_\beta -\Tr\left[{}^1D^{1/2\,1/2}_{1/2\,1/2}\cdot {}^1D^{1/2\,-1/2}_{1/2\,-1/2}  \right] \nonumber\\
&\quad-\Tr\left[{}^1D^{1/2\,1/2}_{1/2\,-1/2}\cdot {}^1D^{1/2\,-1/2}_{1/2\,1/2} \right] \nonumber\\
&\quad +\Tr\left[{}^1D^{1/2\,1/2}_{1/2\,-1/2}\right]\Tr\left[{}^1D^{1/2\,-1/2}_{1/2\,1/2}\right] \,,\nonumber\\
\Tr\left[{}^2\kappa^{00}_{10}\right] &=  \Tr\left[{}^2\kappa^{10}_{00}\right] =0   \,,\nonumber\\
\Tr\left[{}^2\kappa^{11}_{11}\right] &=  N_\alpha^2 - \Tr\left[\left({}^1D^{1/2\,1/2}_{1/2\,1/2}\right)^2\right] \,,\nonumber\\
\Tr\left[{}^2\kappa^{1-1}_{1-1}\right] &=  N_\beta^2 - \Tr\left[\left({}^1D^{1/2\,-1/2}_{1/2\,-1/2}\right)^2\right]   \,.
\end{align}
The third and fourth terms in the trace of ${}^2\kappa^{00}_{00}, {}^2\kappa^{10}_{10}$ vanish for $M$ conserved, as the off-diagonal blocks of the 1RDM coupling $\alpha, \beta$-spins vanish.
As a consistency check, we further observe that the traces of ${}^2\kappa^{\lambda\mu}_{\lambda^\prime\mu^\prime}$ in Eq.~\eqref{eq:traces-kappa} add up to the correct trace of the non-cumulant part of the full 2RDM, ${}^2\kappa\equiv(\mathbbm{1}-\mathbbm{Ex}){}^1D\otimes {}^1D$\,, with \cite{RM15, SHMM18, GBM23}
\begin{equation}
\Tr[{}^2\kappa]=\mathrm{Tr}[{}^1D]^2-\Tr[({}^1D)^2]\,,
\end{equation}
as required.

Eqs.~\eqref{eq:2RDM-adapt} and \eqref{eq:cumulant-2rdm} show that the correct expectation value of the total spin $S$ can be easily incorporated into two-particle RDM methods through linear trace conditions. In addition to the variational 2RDM method \cite{M04, M05, CL06, VAVAB09, NBFFP08, LBSIB15, M20, DLBBKRB23, DP24}, this includes cumulant functional theory \cite{K06, SSS13, SS13, MTSS20} and natural orbital functionals \cite{Piris19, P21}, which approximate the 2RDM in terms of the 1RDM.
Furthermore, the spin-adapted cumulants and the new $S,M$-dependent constraints thereon derived in this section enable a systematic extension of these methods beyond spin-independent Hamiltonians. However, methods operating at the two-particle level cannot fully ensure $S$-representability via linear conditions, as verifying the vanishing variance of $\bd S^2$ requires access to the 4RDM. Therefore, we work out the spin adaptation of the 4RDM and, thus, the corresponding spin-adapted cumulants, in Sec.~\ref{sec:four}.

Independent of the symmetries of a quantum system at hand, the $N$-fermion Hilbert space $\mathcal{H}_N$ can be decomposed into an internal direct sum of sectors labelled by the quantum number $S$ according to $\mathcal{H}_N = \oplus_S \mathcal{H}_N^{(S)}$.
This is especially relevant for Hamiltonians that no longer commute with the $S_z$ operator but still treat the total spin $S$ as an (approximate) good quantum number.
As a consequence of breaking the $\text{SU}(2)$ spin symmetry, the overlap matrix elements $M_{ij}= \langle\Psi|{}^2C^{\lambda\mu}_j({}^2C^{\lambda\mu^\prime}_i)^\dagger|\Psi\rangle$ for the same $\lambda$ but different $\mu, \mu^\prime$ do not vanish anymore. Therefore, the remaining ten spin-adapted blocks of the 2RDM may be nonzero and have to be considered in a variational ground state energy calculation. We provide a list of these additional ten spin-adapted 2RDM matrix elements in Appendix \ref{app:2RDM}. As they are not related to $S,M$ their traces are not constrained in terms of $S,M$ either. Moreover, their cumulants follow from Eq.~\eqref{eq:cumulant-2rdm} analogously to Eq.~\eqref{eq:cadapt-cum}.

\subsection{Spin adaptation on three-particle level \label{sec:three}}

The spin-adapted cumulant of a generic $p$-particle RDM is given by
\begin{equation}\label{eq:cumulant-prdm}
{}^p\Delta^{\lambda\mu}_{\lambda^\prime \mu^\prime}\equiv {}^pD^{\lambda \mu}_{\lambda^\prime \mu^\prime} - {}^p\kappa^{\lambda \mu}_{\lambda^\prime \mu^\prime}\quad \forall \,1\leq p\leq N\,.
\end{equation}
In particular, this defines the 3RDM cumulant ${}^3\Delta$, which is the primary focus of this section.

In contrast to the 1RDM and 2RDM, the spin adaptation of $p$-RDMs with $p\geq 3$ is in general not unique. Previously, the spin adaptation of the 3-RDM has been studied in the context of the partial three-positivity conditions---T1 and T2 conditions~\cite{Mazziotti.2005}. To construct a valid and complete set of spin-adapted three-particle creation operators, we exploit the inverse of Eq.~\eqref{eq:CG-product}, namely
\begin{equation}\label{eq:TO-product}
C^{\lambda, \mu} = \sum_{\mu_1}\begin{bmatrix}
\lambda_1&\lambda_2&\lambda\\ \mu_1& \mu_2&\mu
\end{bmatrix}C^{\lambda_1, \mu_1} C^{\lambda_2, \mu_2}\,.
\end{equation}
Moreover, it is sufficient to construct the irreducible tensor operator element ${}^3C^{\lambda, \lambda}$, as the respective operators for $|\mu|<\lambda$ follow by applying the commutation relations in Eq.~\eqref{eq:tensor-op}.
This eventually leads to the set of operators ${}^3C^{\lambda,\mu}$ presented in Appendix \ref{app:3RDM} (taking also the correct normalization of the 3RDM into account).
Then, the spin-adapted 3RDM matrix elements are obtained analogously to the spin-adapted 2RDM in Eq.~\eqref{eq:2rdm-general}. The resulting spin-adapted 3RDM blocks for $M$ conserved are provided in Appendix \ref{app:3RDM}. The remaining spin-adapted blocks for $M$ not being a good quantum number can be derived from Eq.~\eqref{eq:TO-3rdm} by considering all remaining couplings of two tensor operators from Eq.~\eqref{eq:TO-3rdm}.

The traces of the six non-vanishing spin-adapted 3RDM blocks for $M=0$ are nonzero, while the traces of the four blocks ${}^3D^{3/2\,\pm 1/2}_{1/2\,\pm1/2}, {}^3D^{1/2\,\pm1/2}_{3/2\,\pm 1/2}$ that are possibly nonzero for $M\neq 0$ vanish similar to ${}^2D^{10}_{00}, {}^2D^{00}_{10}$ in Eq.~\eqref{eq:traces}, such that
\begin{align}\label{eq:traces-3RDM}
\Tr\left[{}^3D^{3/2\,3/2}_{3/2\,3/2}\right] &= (N_{\alpha}-2)\Tr\left[{}^2D^{11}_{11}\right]  \,,\nonumber\\
\Tr\left[{}^3D^{3/2\,-3/2}_{3/2\,-3/2}\right] &=  ( N_{\beta}-2)\Tr\left[{}^2D^{1-1}_{1-1}\right]\,,\nonumber\\
\Tr\left[{}^3D^{3/2\,1/2}_{3/2\,1/2} \right]&=\frac{(N_\alpha-1)}{2} \left(3\Tr\left[{}^2D^{10}_{10}\right]-\Tr\left[{}^2D^{00}_{00}\right]\right) \,,\nonumber\\
 \Tr\left[{}^3D^{3/2\,-1/2}_{3/2\,-1/2}\right] &=\frac{(N_\beta-1)}{2} \left(3\Tr\left[{}^2D^{10}_{10}\right]-\Tr\left[{}^2D^{00}_{00}\right]\right)\,,\nonumber\\
\Tr\left[{}^3D^{1/2\,1/2}_{1/2\,1/2}\right] &= 2 (N_\alpha-1)\Tr\left[{}^2D^{00}_{00}\right]\,,\nonumber\\
\Tr\left[{}^3D^{1/2\,-1/2}_{1/2\,-1/2}\right] &= 2(N_\beta-1)\Tr\left[{}^2D^{00}_{00}\right]\,,\nonumber\\
\Tr\left[{}^3D^{3/2\,1/2}_{1/2\,1/2}\right] &=\Tr\left[ {}^3D^{1/2\,1/2}_{3/2\,1/2}\right] =0\,,\nonumber\\
\Tr\left[{}^3D^{3/2\,-1/2}_{1/2\,-1/2}\right]&=\Tr\left[{}^3D^{1/2\,-1/2}_{3/2\,-1/2}\right]= 0   \,.
\end{align}
As a consistency check, we observe that the traces in Eq.~\eqref{eq:traces-3RDM} are in agreement with the normalization $\Tr[{}^3D]=N(N-1)(N-2)$, and with Ref.~\cite{PRTV97}. Next, we use Eq.~\eqref{eq:traces-3RDM} to derive trace conditions for the spin-adapted 3RDM cumulants.

Since the non-cumulant part ${}^3\kappa^{\lambda\mu}_{\lambda^\prime\mu^\prime}$ of the 3RDM is a functional of the 2RDM, Eq.~\eqref{eq:cumulant-prdm} imposes the necessary trace conditions on the 3RDM cumulant ${}^3\Delta^{\lambda\mu}_{\lambda^\prime\mu^\prime}$ for compatibility with an $N$-fermion quantum state of fixed total spin $S$ and magnetization $M$. This eventually leads to (recall that the 1RDM is block-diagonal for $M$ conserved)
\begin{align}\label{eq:kappa-3RDM}
\Tr\left[{}^3\kappa^{3/2\,3/2}_{3/2\,3/2}\right] &=N_{\alpha}^3 - 3N_\alpha^2 -4 \Tr\left[\left({}^1D^{1/2\, 1/2}_{1/2\, 1/2}\right)^3 \right]\nonumber\\
&\quad  + 6 \Tr\left[\left({}^1D^{1/2\, 1/2}_{1/2\, 1/2}\right)^2 \right]\,,\nonumber\\
\Tr\left[{}^3\kappa^{3/2\,-3/2}_{3/2\,-3/2}\right] &=N_{\beta}^3 - 3N_\beta^2  - 4 \Tr\left[\left({}^1D^{1/2\, -1/2}_{1/2 \, -1/2}\right)^3 \right] \nonumber\\
&\quad  +6 \Tr\left[\left({}^1D^{1/2\, -1/2}_{1/2\, -1/2}\right)^2 \right]\,,\nonumber\\
\Tr\left[{}^3\kappa^{3/2\,1/2}_{3/2\,1/2} \right]&=  N_\alpha(N_\alpha N_\beta - N_\beta) - X_\alpha \,,\nonumber\\
 \Tr\left[{}^3\kappa^{3/2\,-1/2}_{3/2\,-1/2}\right] &=   N_\beta( N_\beta N_\alpha - N_\alpha)- X_\beta \,,\nonumber\\
\Tr\left[{}^3\kappa^{1/2\,1/2}_{1/2\,1/2}\right] &=   2N_\alpha \left( N_\alpha N_\beta - N_\beta\right)+X_\alpha\,,\nonumber\\
\Tr\left[{}^3\kappa^{1/2\,-1/2}_{1/2\,-1/2}\right] &=  2N_\beta \left(           N_\alpha N_\beta - N_\alpha\right) +X_\beta   \,,\nonumber\\
\Tr\left[{}^3\kappa^{3/2\,1/2}_{1/2\,1/2}\right] &=\Tr\left[ {}^3\kappa^{1/2\,1/2}_{3/2\,1/2}\right] = 0\,,\nonumber\\
\Tr\left[{}^3\kappa^{3/2\,-1/2}_{1/2\,-1/2}\right]&=\Tr\left[{}^3\kappa^{1/2\,-1/2}_{3/2\,-1/2}\right]= 0   \,,
\end{align}
\begin{widetext}
where $X_\alpha, X_\beta$ are given by
\begin{align}
X_\alpha&\equiv -2 \Tr\left[ {}^3\kappa^{\alpha\beta\alpha}_{\beta\alpha\alpha}\right]=-2 \Tr\left[ {}^3\kappa^{\alpha\alpha\beta}_{\alpha\beta\alpha}\right]   = -2 \Tr\left[ {}^3\kappa^{\alpha\alpha\beta}_{\beta\alpha\alpha}\right] = -2 \Tr\left[ {}^3\kappa^{\alpha\beta\alpha}_{\alpha\alpha\beta}\right]  = -2 \Tr\left[ {}^3\kappa^{\beta\alpha\alpha}_{\alpha\alpha\beta}\right] =-2 \Tr\left[ {}^3\kappa^{\beta\alpha\alpha}_{\alpha\beta\alpha}\right]   \nonumber\\
&=2 \Big[ -N_\alpha\left(S(S+1)-M^2-\frac{N}{2}\right) - \Tr\left[{}^1D^{1/2\, 1/2}_{1/2\, 1/2} \cdot {}^1D^{1/2\, -1/2}_{1/2\, -1/2}\right] + 2\Tr\left[\left({}^1D^{1/2\, 1/2}_{1/2\, 1/2}\right)^2\cdot {}^1D^{1/2\, -1/2}_{1/2\, -1/2}\right]\nonumber\\
&\quad + \sum_{p,q,r=1}^d  \left({}^1D^{1/2\, 1/2}_{1/2\, 1/2}\right)^r_q \left({}^2D^{p\beta, q\alpha}_{p\alpha, r\beta} + {}^2D^{p\alpha, q\beta}_{p\beta, r\alpha} \right)\Big]\,,\nonumber\\
X_\beta &\equiv -2 \Tr\left[ {}^3\kappa^{\beta\alpha\beta}_{\alpha\beta\beta}\right] = -2 \Tr\left[ {}^3\kappa^{\beta\alpha\beta}_{\alpha\beta\beta}\right] =-2 \Tr\left[ {}^3\kappa^{\beta\beta\alpha}_{\beta\alpha\beta}\right] =-2 \Tr\left[ {}^3\kappa^{\beta\beta\alpha}_{\alpha\beta\beta}\right]=-2 \Tr\left[ {}^3\kappa^{\alpha\beta\beta}_{\beta\alpha\beta}\right] =-2 \Tr\left[ {}^3\kappa^{\alpha\beta\beta}_{\beta\beta\alpha}\right] \nonumber\\
& =2 \Big[ -N_\beta\left(S(S+1)-M^2-\frac{N}{2}\right) - \Tr\left[{}^1D^{1/2\, -1/2}_{1/2\, -1/2} \cdot {}^1D^{1/2\, 1/2}_{1/2\, 1/2}\right] + 2\Tr\left[\left({}^1D^{1/2\, -1/2}_{1/2\, -1/2}\right)^2\cdot {}^1D^{1/2\, 1/2}_{1/2\, 1/2}\right]\nonumber\\
&\quad + \sum_{p,q,r=1}^d  \left({}^1D^{1/2\, -1/2}_{1/2\, -1/2}\right)^r_q \left({}^2D^{p\beta, q\alpha}_{p\alpha, r\beta} + {}^2D^{p\alpha, q\beta}_{p\beta, r\alpha} \right)\Big]\,,
\end{align}
\end{widetext}
and ${}^2D^{p\sigma, q\sigma^\prime}_{p\sigma^\prime, r\sigma}$ with $\sigma, \sigma^\prime\in \{\alpha, \beta\}$ can be expressed in terms of the spin-adapted 2RDM blocks ${}^2D^{\lambda\mu}_{\lambda^\prime\mu^\prime}$ according to Eq.~\eqref{app:2RDM-vs-adapted}. In particular, Eq.~\eqref{eq:kappa-3RDM} ensures the correct trace for the three-particle non-cumulant term ${}^3\kappa$:
\begin{equation}
\Tr\left[{}^3\kappa\right] = N^3 -  3 N^2 - 4 \Tr\Big[\left({}^1D\right)^3 \Big]+6\Tr\Big[\left({}^1D\right)^2\Big]\,.
\end{equation}

\subsection{Ensuring $S$-representability on the four-particle level \label{sec:four}}

As noted in Sec.~\ref{sec:two}, the trace conditions on the 2RDM in Eq.~\eqref{eq:traces} ensure the correct expectation value of $\bd S^2$. However, RDM theories limited to $p$-RDMs with $p\leq 3$ cannot enforce $S$-representability solely through such trace constraints. Instead, deriving complex $S$-representability conditions for admissible $p$-RDMs would be required.
To avoid these highly complicated constraints, we first observe that for any $N$-fermion state with well-defined total spin $S$, i.e., $\ket{\Psi}\in \mathcal{H}_N^{(S)}$ with $\mathcal{H}_N^{(S)}$ as introduced in Sec.~\ref{sec:two}, the variance $\Delta\bd S^2$ of the $\bd S^2$ operator satisfies
\begin{equation}\label{eq:variance-S}
\Delta\bd S^2 =\bra{\Psi}\left(\bd S^2-\langle\bd S^2\rangle\right)^2\ket{\Psi} = \langle\bd S^4\rangle -\langle\bd S^2\rangle^2=0\,.
\end{equation}

The variance of $\bd S^2$ in Eq.~\eqref{eq:variance-S} directly reduces to a linear functional of the 4RDM. We now recall Rosina's theorem \cite{Rosina68,M98-cse} which states that for Hamiltonians with at most two-body interactions and a non-degenerate ground state, there is an one-to-one relation between the ground state 2RDM and the wave function as well as higher $p$-RDMs. Furthermore, Ref.~\cite{M98-cse} demonstrated that this bijection between 2RDMs and $p$-RDMs with $3\leq p\leq N$ also holds in the case of degenerate pure states that can be distinguished by a two-particle operator. This is relevant for spin symmetries, as the states of a spin multiplet with the same $S$ but different $M$ can be distinguished by (the square of the) $S_z$ operator.
Thus, it follows that vanishing of the variance $\Delta\bd S^2$ is indeed a necessary and sufficient condition for an 4RDM to be compatible with an $N$-fermion quantum state with total spin $S$.

The operator $\bd S^4$ and, thus, also the variance $\Delta\bd S^2$ can be expressed in terms of the spin-adapted 4RDM matrix elements leading to conditions on their traces in terms of expectation values of the spin operators. The spin-adapted 4RDM blocks follow from the spin-adapted four-particle creation operators ${}^4C^{\lambda, \mu}$ as for the lower RDMs in Secs.~\ref{sec:one}-\ref{sec:three}. Since four spin-$1/2$ particles can be coupled to a quintet, triplet, or a singlet, there are in total nine irreducible tensor operator components ${}^4C^{\lambda, \mu}$. A valid set of respective tensor operators is conveniently constructed using Eqs.~\eqref{eq:tensor-op} and \eqref{eq:TO-product}, and we provide their explicit expressions in Appendix \ref{app:4rdm}. The spin-adapted 4RDM matrix elements follow accordingly, and their cumulants are defined through Eq.~\eqref{eq:cumulant-prdm}. In particular, the trace conditions restricting the set of $S$-representable 4RDM cumulants follow from the traces of ${}^4\kappa^{\lambda\mu}_{\lambda^\prime\mu^\prime}$ and the spin-adapted 4RDM trace conditions.

\section{Symmetry adaptation in the presence of spin-orbit coupling \label{sec:LS}}

In this section, we extend the spin adaptation of RDMs and their cumulants, developed in Sec.~\ref{sec:theory}, to the case of spin-orbit coupling, which becomes significant for heavy atoms and compounds where relativistic effects are non-negligible \cite{R07, Reiher14}. We consider a local spin-orbit interaction, where the orbital degrees of freedom are coupled to the spin of the same electron, neglecting spin-other-orbit coupling and spin-spin couplings as higher-order corrections to the non-relativistic electronic structure Hamiltonian \cite{GB93-so, Moshinsky69}. The spin-orbit coupling is described by the one-particle Hamiltonian $h_{so} = \sum_{i=1}^N \bd L_i\cdot\bd S_i$, where $\bd L_i, \bd S_i$ are the orbital angular momentum and spin operators for electron $i\in\{1, ..., N\}$. Although $S_z, L_z$ are not conserved individually, a set of mutually commuting symmetry operators is given by $\bd L^2, \bd S^2, \bd J^2, J_z$ \cite{Moshinsky69}. Thus, the total angular momentum $\bd J = \bd L+\bd S$ and its projection $J_z$ onto the magnetization axis are good quantum numbers. Therefore, it is convenient to introduce the symmetry adaptation of the RDMs and their cumulants in the total angular momentum representation.

We begin by introducing the irreducible tensor operators in the angular momentum representation on the one-particle level. The total angular momentum quantum number is denoted by $J$, and the eigenvalue of the $J_z$ operator is denoted by $M_j$ to distinguish it from $M$ defined in Sec.~\ref{sec:one}, as well as from the eigenvalue $M_l$ of $L_z$. Following standard references \cite{Moshinsky69}, we denote the fermionic creation and annihilation operators corresponding to an eigenstate of $\bd L^2, L_z, \bd S^2, S_z$ by $f_{\xi, L, M_l, M}^\dagger$ and $f_{\xi, L, M_l, M}$, where $M=\pm 1/2$ denotes the eigenvalue of $S_z$ and $\xi$ represents a set of quantum numbers and labels characterizing the one-particle state. For instance, $\xi = r$ in coordinate space or $\xi=n$, where $n$ is the principal quantum number in the case of atoms. Additionally, the operators $f_{\xi, L, M_l, M_s}^\dagger$ are themselves tensor operators.

Since $\bd J$ describes the coupling of the orbital angular momentum $\bd L$ and the spin angular momentum $\bd S$, it follows that $J=L\pm 1/2$. Therefore, combining Eq.~\eqref{eq:TO-product} with the Clebsch-Gordon coefficients for coupling of $\bd L$ with $\bd S$ of a spin-$1/2$ particle (see Ref.~\cite{khersonskii1988quantum})
yields the fermionic creation/annihilation operators $f_{\xi, J=L\pm 1/2, M_j}^\dagger/f_{\xi, J=L\pm 1/2, M_j}$ in total angular momentum representation.
These fermionic creation/annihilation operators are components of irreducible tensor operators similar to those in Sec.~\ref{sec:one}. The range of $J=L\pm 1/2$ depends on the possible values of $L$ and is therefore system-specific. For example, if $\xi=n$ is the principal quantum number, then $L\in \{0, 1, ..., n-1\}$, which determines the number of total angular momentum adapted 1RDM blocks.


Based on the total angular momentum representation described above, the symmetry-adapted RDM matrix elements for $p$-RDMs and their cumulants can be defined and calculated by following the same steps as in Secs.~\ref{sec:two}-\ref{sec:four}. Since the possible values for the total angular momentum $J$ for $p \geq 2$ depend on $L$ at the one-particle level and are thus system dependent, we do not provide a specific example here. Furthermore, this adaptation of the RDM matrix elements in the angular momentum representation enables the identification of the non-vanishing symmetry-adapted $p$-RDM blocks, analogous to Secs.~\ref{sec:two}-\ref{sec:four}.

Then, $J$-representability of $p$-RDMs is defined in analogy to $S$-representability in Sec.~\ref{sec:theory}.
As $\bd J^2$ is again a two-particle operator, a necessary and sufficient condition for $J$-representability is given by a vanishing variance of $\bd J^2$ for $N$-representable 4RDMs as discussed in detail for $\bd S^2$ in Sec.~\ref{sec:four}.

\section{Conclusions and outlook}

We proposed and developed the spin adaptation of $p$-RDM cumulants and explored the incorporation of spin symmetries through linear constraints in functional theories at the $p$-particle level. Specifically, we derived new trace conditions for the spin-adapted 2RDM and 3RDM cumulants, which are necessary constraints on the cumulants in terms of the quantum numbers $S$ and $M$.

Our results provide the framework for incorporating spin symmetries in various density matrix functional theories operating on the one-particle to the four-particle level. While lower-level functional theories such as DFT, RDMFT, and 2RDM methods require an explicit $S$ dependence of the universal functional, which is difficult to model in practice, $S$-representability at the four-particle level is ensured by the vanishing variance of the Casimir operator $\bd S^2$.


The spin adaptation of RDM cumulants developed in this work can enhance time-independent methods that use cumulant expansions, including parametric 2-RDM theory~\cite{DM07, DKM08, M08, DM09, M10, DM10-1, DM10-2, SDM11} and the contracted Schrödinger equation~\cite{CV93, YN97, M98-cse, M99-cse, AV01, HH02, MK01, M06, M07-acse, M07-acse-jcp, M07-acse-pra, RFM09, GM09, AVTPRO11, SM15, SM21}, by providing a systematic framework for the spin adaptation of 3- and 4-particle cumulants. Spin-adapted 3-RDM reconstructions also play a central role in time-dependent 2-RDM methods~\cite{LBSIB15, LBTIB17, DLBBKRB23, BSDR24, PEB24}. In addition, the present framework extends naturally to spin-dependent Hamiltonians, enabling the analysis of contributions from individual spin-adapted blocks to the ground-state energy and other observables derived from reduced density matrices.

Additionally, symmetry adaptation in the total angular momentum representation can be used to identify nonzero $p$-RDM blocks and derive corresponding trace conditions on their cumulants, as discussed in Sec.~\ref{sec:LS}. Beyond these applications, spin-adapted cumulants and their trace conditions can be utilized in functional theories such as density cumulant theory \cite{K06, SSS13, SS13, CS18, MTSS20} or natural orbital functional theory \cite{Piris19, Piris21-1, P21}.

Finally, we would like to stress the generality of the symmetry adaptation of $p$-RDMs and their cumulants presented in this work. In fact, this framework can be extended to other symmetries described by Lie groups \cite{Miller72-book, Hall15}. Generally, this involves constructing irreducible tensor operators for an irreducible unitary representation of the given group. The derivation of symmetry-adapted $p$-RDMs then depends on the ability to define symmetry-adapted $p$-particle creation operators.
Once the symmetry-adapted $p$-RDMs are constructed, their cumulants and non-cumulant parts can be derived similarly to Eq.~\eqref{eq:cumulant-prdm}.

\begin{acknowledgments}
We acknowledge financial support from the U.S. National Science Foundation (NSF) Grant No. CHE-2155082 (D.M.), the German Research Foundation (Grant SCHI 1476/1-1) (J.L., C.S.), the Munich Center for Quantum Science and Technology and the International Max Planck Research School for Quantum Science and Technology (IMPRS-QST) (J.L.). The project/research is also part of the Munich Quantum Valley, which is supported by the Bavarian state government with funds from the Hightech Agenda Bayern Plus.
\end{acknowledgments}

\appendix
\onecolumngrid

\section{Spin-adapted 2RDM matrix elements \label{app:2RDM}}

In this section, we present the spin-adapted 2RDM matrix elements and the additional spin-adapted 2RDM blocks, which vanish when $M$ is conserved but must be considered otherwise.

Inverting the expressions for ${}^2D^{\lambda\mu}_{\lambda\mu}$ in Eq.~\eqref{eq:2RDM-adapt} for the spin-$\alpha, \beta$ blocks of the 2RDM yields
\begin{eqnarray}\label{app:2RDM-vs-adapted}
^2D^{i\beta, j\alpha}_{k\beta, l\alpha} &=& \frac{1}{2}\left[\left( ^2D^{00}_{00}\right)^{ij}_{kl}+\left( ^2D^{10}_{10}\right)^{ij}_{kl} +\left( ^2D^{00}_{10}\right)^{ij}_{kl} +\left( ^2D^{10}_{00}\right)^{ij}_{kl}\right],\nonumber\\
^2D^{i\alpha, j\beta}_{k\alpha, l\beta} &=& \frac{1}{2}\left[\left( ^2D^{00}_{00}\right)^{ij}_{kl}+\left( ^2D^{10}_{10}\right)^{ij}_{kl} -\left( ^2D^{00}_{10}\right)^{ij}_{kl} -\left( ^2D^{10}_{00}\right)^{ij}_{kl}\right],\nonumber\\
^2D^{i\alpha, j\beta}_{k\beta, l\alpha} &=& \frac{1}{2}\left[\left( ^2D^{10}_{10}\right)^{ij}_{kl} +\left( ^2D^{10}_{00}\right)^{ij}_{kl} -\left( ^2D^{00}_{10}\right)^{ij}_{kl} -\left( ^2D^{00}_{00}\right)^{ij}_{kl}\right],\nonumber\\
^2D^{i\beta, j\alpha}_{k\alpha, l\beta} &=& \frac{1}{2}\left[\left( ^2D^{10}_{10}\right)^{ij}_{kl} +\left( ^2D^{00}_{10}\right)^{ij}_{kl} -\left( ^2D^{10}_{00}\right)^{ij}_{kl} -\left( ^2D^{00}_{00}\right)^{ij}_{kl}\right],\nonumber\\
^2D^{i\alpha, j\alpha}_{k\alpha, l\alpha} &=& \left( ^2D^{11}_{11}\right)^{ij}_{kl},\nonumber\\
^2D^{i\beta, j\beta}_{k\beta, l\beta} &=& \left( ^2D^{1-1}_{1-1}\right)^{ij}_{kl}\,.
\end{eqnarray}
%

For non-conserved $M$, the matrix elements of the ten spin-adapted 2RDM blocks complementing Eq.~\eqref{eq:2RDM-adapt} follow from the spin-adapted two-particle creation operators in Eq.~\eqref{eq:tos-2rdm} as
\begin{align}\label{eq:add-blocks}
\left({}^2D^{10}_{11}\right)^{ij}_{kl} &= \frac{1}{\sqrt{2}}\left( {}^2D^{i\alpha,j\beta}_{k\alpha,l\alpha} +{}^2D^{i\beta,j\alpha}_{k\alpha,l\alpha}\right),\nonumber \\
\left({}^2D^{11}_{10}\right)^{ij}_{kl} &= \frac{1}{\sqrt{2}}\left( {}^2D^{i\alpha,j\alpha}_{k\alpha,l\beta} +{}^2D^{i\alpha,j\alpha}_{k\beta,l\alpha}\right),\nonumber \\
\left({}^2D^{10}_{1-1}\right)^{ij}_{kl} &= \frac{1}{\sqrt{2}}\left( {}^2D^{i\alpha,j\beta}_{k\beta,l\beta} +{}^2D^{i\beta,j\alpha}_{k\beta,l\beta}\right) ,\nonumber \\
\left({}^2D^{1-1}_{10}\right)^{ij}_{kl} &= \frac{1}{\sqrt{2}}\left( {}^2D^{i\beta,j\beta}_{k\alpha,l\beta} +{}^2D^{i\beta,j\beta}_{k\beta,l\alpha}\right) ,\nonumber \\
\left({}^2D^{00}_{11}\right)^{ij}_{kl} &= \frac{1}{\sqrt{2}}\left( {}^2D^{i\beta,j\alpha}_{k\alpha,l\alpha} -{}^2D^{i\alpha,j\beta}_{k\alpha,l\alpha}\right) ,\nonumber \\
\left({}^2D^{11}_{00}\right)^{ij}_{kl} &=  \frac{1}{\sqrt{2}}\left( {}^2D^{i\alpha,j\alpha}_{k\beta,l\alpha} -{}^2D^{i\alpha,j\alpha}_{k\alpha,l\beta}\right),\nonumber \\
\left({}^2D^{00}_{1-1}\right)^{ij}_{kl} &= \frac{1}{\sqrt{2}}\left( {}^2D^{i\beta,j\alpha}_{k\beta,l\beta} -{}^2D^{i\alpha,j\beta}_{k\beta,l\beta}\right) ,\nonumber \\
\left({}^2D^{1-1}_{00}\right)^{ij}_{kl} &= \frac{1}{\sqrt{2}}\left( {}^2D^{i\beta,j\beta}_{k\beta,l\alpha} -{}^2D^{i\beta,j\beta}_{k\alpha,l\beta}\right),\nonumber \\
\left({}^2D^{11}_{1-1}\right)^{ij}_{kl} &= {}^2D^{i\alpha,j\alpha}_{k\beta, l\beta},\nonumber\\
\left({}^2D^{1-1}_{11}\right)^{ij}_{kl} &= {}^2D^{i\beta,j\beta}_{k\alpha, l\alpha}\,.
\end{align}
Their spin-adapted cumulants and non-cumulant parts follow from Eq.~\eqref{eq:cumulant-2rdm}.

\section{Spin-adapted 3RDM matrix elements \label{app:3RDM}}

In this section, we derive the irreducible tensor operators on the three-particle level and work out the spin adaptation of the 3RDM.
Since three spin-$1/2$ fermions can couple to doublet or quartet, there are six irreducible tensor operators.
These irreducible tensor operators for the spin adaptation of the 3RDM follow from Eq.~\eqref{eq:TO-product} as
\begin{align}\label{eq:TO-3rdm}
{}^3 C^{3/2,3/2}_{kji} &\equiv{}^2 C^{1,1}_{ji}\,\, {}^1C^{1/2,1/2}_k= f_{i\alpha}^\dagger f_{j\alpha}^\dagger  f_{k\alpha}^\dagger\,,\nonumber\\
C^{3/2,1/2}_{kji}&\equiv \frac{1}{\sqrt{3}}\left(\sqrt{2}  {}^2C^{1,0}_{ji}\,\, {}^1C^{1/2,1/2}_k+ {}^2C^{1,1}_{ji} \,\, {}^1C^{1/2,-1/2}_k\right)=\frac{1}{\sqrt{3}}\left(f_{i\beta}^\dagger f_{j\alpha}^\dagger  f_{k\alpha}^\dagger  +f_{i\alpha}^\dagger f_{j\beta}^\dagger  f_{k\alpha}^\dagger+f_{i\alpha}^\dagger f_{j\alpha}^\dagger  f_{k\beta}^\dagger\right) \,,\nonumber\\
{}^3C^{3/2,-1/2}_{kji} &\equiv \frac{1}{\sqrt{3}} \left(\sqrt{2} {}^2C^{1,0}_{ji}\,\, {}^1C^{1/2,-1/2}_k+ {}^2C^{1,-1}_{ji}\,\, {}^1C^{1/2,1/2}_k\right)=\frac{1}{\sqrt{3}}\left(f_{i\beta}^\dagger f_{j\alpha}^\dagger  f_{k\beta}^\dagger  +f_{i\alpha}^\dagger f_{j\beta}^\dagger  f_{k\beta}^\dagger+f_{i\beta}^\dagger f_{j\beta}^\dagger  f_{k\alpha}^\dagger\right) \,,\nonumber\\
{}^3C^{3/2,-3/2}_{kji} &\equiv{}^2C^{1,-1}_{ji} \,\, {}^1C^{1/2,-1/2}_k= f_{i\beta}^\dagger f_{j\beta}^\dagger  f_{k\beta}^\dagger\,,\nonumber\\
{}^3C^{1/2,1/2}_{kji} &\equiv\sqrt{2} {}^2C^{0,0}_{ji}\,\, {}^1C^{1/2,1/2}_k=\left(f_{k\alpha}^\dagger f_{l\beta}^\dagger f_{m\alpha}^\dagger - f_{k\beta}^\dagger f_{l\alpha}^\dagger f_{m\alpha}^\dagger  \right)\,,\nonumber\\
{}^3 C^{1/2,-1/2} &\equiv \sqrt{2}{}^2C^{0,0}_{ji} \,\, {}^1C^{1/2,-1/2}_k =\left(f_{k\alpha}^\dagger f_{l\beta}^\dagger f_{m\beta}^\dagger - f_{k\beta}^\dagger f_{l\alpha}^\dagger f_{m\beta}^\dagger  \right)\,.
\end{align}
Therefore, we obtain the following set of spin-adapted 3RDM blocks for $M$ conserved:
\begin{align}
{}^3D^{3/2\,3/2}_{3/2\,3/2} &= {}^3D^{\alpha \alpha \alpha}_{\alpha\alpha\alpha}\,,\nonumber\\
{}^3D^{3/2\,1/2}_{3/2\,1/2} &= \frac{1}{3}\left({}^3D^{\alpha \alpha \beta}_{\alpha\alpha\beta}+ {}^3D^{\alpha \alpha \beta}_{\alpha\beta\alpha}+{}^3D^{\alpha \alpha \beta}_{\beta\alpha\alpha}+ {}^3D^{\alpha\beta\alpha}_{\alpha\alpha\beta}+ {}^3D^{\alpha\beta\alpha}_{\alpha\beta\alpha}+{}^3D^{\alpha\beta\alpha}_{\beta\alpha\alpha}+{}^3D^{\beta\alpha\alpha}_{\alpha\alpha\beta}+ {}^3D^{\beta\alpha\alpha}_{\alpha\beta\alpha}+{}^3D^{\beta\alpha\alpha}_{\beta\alpha\alpha}
  \right)\,, \nonumber\\
{}^3D^{3/2\,1/2}_{1/2\,1/2} &= \frac{1}{\sqrt{3}}\left({}^3D^{\alpha\alpha\beta}_{\alpha\beta\alpha} - {}^3D^{\alpha\alpha\beta}_{\alpha\alpha\beta}
+{}^3D^{\alpha\beta\alpha}_{\alpha\beta\alpha} - {}^3D^{\alpha\beta\alpha}_{\alpha\alpha\beta} +{}^3D^{\beta\alpha\alpha}_{\alpha\beta\alpha} - {}^3D^{\beta\alpha\alpha}_{\alpha\alpha\beta}
 \right) \,,\nonumber\\
 {}^3D^{1/2\,1/2}_{3/2\,1/2} &= \frac{1}{\sqrt{3}}\left({}^3D_{\alpha\alpha\beta}^{\alpha\beta\alpha} - {}^3D_{\alpha\alpha\beta}^{\alpha\alpha\beta}
+{}^3D_{\alpha\beta\alpha}^{\alpha\beta\alpha} - {}^3D_{\alpha\beta\alpha}^{\alpha\alpha\beta} +{}^3D_{\beta\alpha\alpha}^{\alpha\beta\alpha} - {}^3D_{\beta\alpha\alpha}^{\alpha\alpha\beta}
 \right) \,,\nonumber\\
 {}^3D^{3/2\,-1/2}_{3/2\,-1/2} &=  \frac{1}{3}\left( {}^3D^{\beta\alpha \beta}_{\beta\alpha\beta}+{}^3D^{\beta\alpha \beta}_{\beta\beta\alpha}+{}^3D^{\beta\alpha \beta}_{\alpha\beta\beta} + {}^3D^{\beta\beta\alpha}_{\beta\alpha\beta}+{}^3D^{\beta\beta\alpha}_{\beta\beta\alpha}+{}^3D^{\beta\beta\alpha}_{\alpha\beta\beta}
 + {}^3D^{\alpha\beta\beta}_{\beta\alpha\beta}+{}^3D^{\alpha\beta\beta}_{\beta\beta\alpha}+{}^3D^{\alpha\beta\beta}_{\alpha\beta\beta}
     \right)  \,,  \nonumber\\
{}^3D^{3/2\,-1/2}_{1/2\,-1/2} &=  \frac{1}{\sqrt{3}}\left({}^3D^{\beta \alpha \beta}_{\beta\beta\alpha}- {}^3D^{\beta \alpha \beta}_{\beta\alpha\beta}
+{}^3D^{\beta \beta\alpha}_{\beta\beta\alpha}- {}^3D^{\beta \beta\alpha}_{\beta\alpha\beta}+{}^3D^{\alpha\beta \beta}_{\beta\beta\alpha}- {}^3D^{\alpha\beta \beta}_{\beta\alpha\beta}
\right) \,,\nonumber\\
{}^3D^{1/2\,-1/2}_{3/2\,-1/2} &=  \frac{1}{\sqrt{3}}\left({}^3D_{\beta \alpha \beta}^{\beta\beta\alpha}- {}^3D_{\beta \alpha \beta}^{\beta\alpha\beta}
+{}^3D_{\beta \beta\alpha}^{\beta\beta\alpha}- {}^3D_{\beta \beta\alpha}^{\beta\alpha\beta}+{}^3D_{\alpha\beta \beta}^{\beta\beta\alpha}- {}^3D_{\alpha\beta \beta}^{\beta\alpha\beta}
\right) \,,\nonumber\\
{}^3D^{3/2\,-3/2}_{3/2\,-3/2} &=  {}^3D^{\beta\beta\beta}_{\beta\beta\beta} \,,\nonumber\\
{}^3D^{1/2\,1/2}_{1/2\,1/2} &= \left({}^3D^{\alpha \beta \alpha}_{\alpha\beta\alpha} - {}^3D^{\alpha \beta \alpha}_{\alpha\alpha\beta}  - {}^3D^{\alpha\alpha\beta}_{\alpha\beta\alpha}+ {}^3D^{\alpha \alpha\beta}_{\alpha\alpha\beta}\right)\,,\nonumber\\
{}^3D^{1/2\,-1/2}_{1/2\,-1/2} &= \left({}^3D^{\beta\beta\alpha}_{\beta\beta\alpha}- {}^3D^{\beta\beta\alpha}_{\beta\alpha\beta} - {}^3D^{\beta\alpha\beta}_{\beta\beta\alpha} + {}^3D^{\beta\alpha\beta}_{\beta\alpha\beta} \right) \,.\nonumber\\
\end{align}

\section{Spin-adapted tensor operators for the 4RDM \label{app:4rdm}}

In this section, we construct the irreducible tensor operators for the spin-adapted 4RDM blocks. Since four particles can couple either to a singlet, triplet, or quintet, we obtain in total nine different tensor operators ${}^4C^{\lambda\mu}$.

We begin by deriving a set of quintet tensor operators for the spin-adapted 4RDM matrix elements based on Eq.~\eqref{eq:TO-product}, leading to
\begin{align}
{}^4 C^{2,2}_{lkji} &= \left(C^{1,1}C^{1,1}\right)_{lkji}\nonumber\\
&= f_{i\alpha}^\dagger f_{j\alpha}^\dagger  f_{k\alpha}^\dagger f_{l\alpha}^\dagger\\
{}^4C^{2,1}_{lkji} &= \frac{1}{\sqrt{2}}\left(C^{1,0}C^{1,1}+C^{1,1}C^{1,0}\right)_{lkji}\nonumber\\
&=\frac{1}{2}\left( f_{i\beta}^\dagger f_{j\alpha}^\dagger  f_{k\alpha}^\dagger f_{l\alpha}^\dagger + f_{i\alpha}^\dagger f_{j\beta}^\dagger  f_{k\alpha}^\dagger f_{l\alpha}^\dagger + f_{i\alpha}^\dagger f_{j\alpha}^\dagger  f_{k\beta}^\dagger f_{l\alpha}^\dagger + f_{i\alpha}^\dagger f_{j\alpha}^\dagger  f_{k\alpha}^\dagger f_{l\beta}^\dagger\right)\,,\nonumber\\
{}^4  C^{2,0}_{lkji} &= \left(\frac{1}{\sqrt{6}}C^{1,1}C^{1,-1}+\frac{1}{\sqrt{6}}C^{1,-1}C^{1,1}+ \sqrt{\frac{2}{3}} C^{1,0}C^{1,0}\right)_{lkji}\nonumber\\
&=\frac{1}{\sqrt{6}}\left(f_{i\beta}^\dagger f_{j\beta}^\dagger  f_{k\alpha}^\dagger f_{l\alpha}^\dagger + f_{i\alpha}^\dagger f_{j\beta}^\dagger  f_{k\beta}^\dagger f_{l\alpha}^\dagger +f_{i\alpha}^\dagger f_{j\beta}^\dagger  f_{k\alpha}^\dagger f_{l\beta}^\dagger+f_{i\beta}^\dagger f_{j\alpha}^\dagger  f_{k\beta}^\dagger f_{l\alpha}^\dagger +f_{i\beta}^\dagger f_{j\alpha}^\dagger  f_{k\alpha}^\dagger f_{l\beta}^\dagger+f_{i\alpha}^\dagger f_{j\alpha}^\dagger  f_{k\beta}^\dagger f_{l\beta}^\dagger \right)\,,\nonumber\\
{}^4C^{2,-1}_{lkji}&=  \frac{1}{\sqrt{2}}\left(C^{1,0}C^{1,-1}+C^{1,-1}C^{1,0}\right)_{lkji}\nonumber\\
&= \frac{1}{2}\left( f_{i\beta}^\dagger f_{j\beta}^\dagger  f_{k\alpha}^\dagger f_{l\beta}^\dagger + f_{i\beta}^\dagger f_{j\beta}^\dagger  f_{k\beta}^\dagger f_{l\alpha}^\dagger + f_{i\alpha}^\dagger f_{j\beta}^\dagger  f_{k\beta}^\dagger f_{l\beta}^\dagger + f_{i\beta}^\dagger f_{j\alpha}^\dagger  f_{k\beta}^\dagger f_{l\beta}^\dagger\right)\,,\nonumber\\
{}^4C^{2,-2}_{lkji} &=\left(C^{1,-1}C^{1,-1}\right)_{lkji}\nonumber\\
&= f_{i\beta}^\dagger f_{j\beta}^\dagger  f_{k\beta}^\dagger f_{l\beta}^\dagger\,.
\end{align}

The triplet operators follow as
\begin{align}
{}^4C^{1,1}_{lkji} &= \sqrt{3} \left(C^{1,1}C^{0,0}\right)_{lkji}=\sqrt{\frac{3}{2}}\left(f_{i\alpha}^\dagger f_{j\alpha}^\dagger f_{k\alpha}^\dagger f_{l\beta}^\dagger -f_{i\alpha}^\dagger f_{j\alpha}^\dagger f_{k\beta}^\dagger f_{l\alpha}^\dagger\right)\,,\nonumber\\
{}^4C^{1,0}_{lkji}&= \sqrt{3} \left(C^{1,0}C^{0,0}\right)_{lkji}=\frac{\sqrt{3}}{2}\left(f_{i\alpha}^\dagger f_{j\beta}^\dagger + f_{i\beta}^\dagger f_{j\alpha}^\dagger\right)\left(f_{k\alpha}^\dagger f_{l\beta}^\dagger - f_{k\beta}^\dagger f_{l\alpha}^\dagger\right)\,,\nonumber\\
{}^4C^{1,-1}_{lkji} &= \sqrt{3}\left(C^{1,-1}C^{0,0}\right)_{lkji}=\sqrt{\frac{3}{2}}\left(f_{i\beta}^\dagger f_{j\beta}^\dagger f_{k\alpha}^\dagger f_{l\beta}^\dagger -f_{i\beta}^\dagger f_{j\beta}^\dagger f_{k\beta}^\dagger f_{l\alpha}^\dagger\right)
\end{align}
and the singlet operator is given by
\begin{align}
{}^4C^{0,0}_{lkji} &= \sqrt{2}\left(C^{0,0}C^{0,0}\right)_{lkji}= \frac{1}{\sqrt{2}}\left(f_{i\alpha}^\dagger f_{j\beta}^\dagger - f_{i\beta}^\dagger f_{j\alpha}^\dagger\right)\left(f_{k\alpha}^\dagger f_{l\beta}^\dagger - f_{k\beta}^\dagger f_{l\alpha}^\dagger\right)\,.
\end{align}

As with the 2RDM and 3RDM, we can use the Wigner-Eckart theorem, which implies that only the spin-adapted 4RDM matrix elements coupling the same $M$ values in ${}^4D^{S,M}_{S^\prime, M^\prime}=\langle {}^4C^{S,M}({}^4C^{S^\prime, M^\prime})^\dagger\rangle $ are nonzero when $M$ is conserved. This results into the following blocks on the diagonal of the 4RDM:
\begin{align}
{}^4 D^{22}_{2 2} &= {}^4D^{\alpha\alpha \alpha \alpha}_{\alpha\alpha \alpha \alpha} \nonumber\\
{}^4 D^{21}_{2 1} &=\frac{1}{4}\left( {}^4D^{\alpha\alpha \alpha \beta}_{\alpha\alpha \alpha \beta} + {}^4D^{\alpha\alpha \alpha \beta}_{\alpha\alpha \beta \alpha} + {}^4D^{\alpha\alpha \alpha \beta}_{\alpha\beta \alpha \alpha} + {}^4D^{\alpha\alpha \alpha \beta}_{\beta\alpha \alpha \alpha}
+  {}^4D^{\alpha \alpha \beta \alpha}_{\alpha\alpha \beta \alpha}  + {}^4D^{\alpha \alpha \beta \alpha}_{\alpha\beta\alpha \alpha} + {}^4D^{\alpha \alpha \beta \alpha}_{\beta\alpha \alpha \alpha}+ {}^4D^{\alpha \alpha \beta \alpha}_{\alpha\alpha \alpha \beta}\right.\nonumber\\
&\quad \quad \left.  +{}^4D^{\alpha \beta \alpha \alpha}_{\alpha\beta \alpha \alpha}+ {}^4D^{\alpha \beta \alpha \alpha}_{\alpha\alpha \alpha \beta} + {}^4D^{\alpha \beta \alpha \alpha}_{\alpha\alpha \beta \alpha}  +{}^4D^{\alpha \beta \alpha \alpha}_{\beta\alpha \alpha \alpha}  +   {}^4D^{\beta \alpha \alpha \alpha}_{\alpha\beta \alpha \alpha}+ {}^4D^{\beta\alpha \alpha \alpha}_{\alpha\alpha \alpha \beta} + {}^4D^{\beta \alpha \alpha \alpha}_{\alpha\alpha \beta \alpha}  +{}^4D^{\beta \alpha \alpha \alpha}_{\beta\alpha \alpha \alpha} \right)\nonumber\\
{}^4 D^{11}_{1 1} &=\frac{3}{2}\left({}^4D_{\beta\alpha \alpha \alpha}^{\beta\alpha \alpha \alpha} -{}^4D_{\beta\alpha \alpha \alpha}^{\alpha\beta \alpha \alpha} +{}^4D_{\alpha\beta \alpha \alpha}^{\alpha \beta \alpha \alpha} - {}^4D_{\alpha\beta\alpha \alpha}^{\beta\alpha \alpha \alpha}     \right)   \nonumber\\
{}^4 D^{2 0}_{2 0} &=\frac{1}{6}\left({}^4D^{\alpha\alpha \beta \beta}_{\alpha\alpha \beta \beta}+{}^4D^{\alpha\alpha \beta \beta}_{\alpha\beta \beta \alpha}+{}^4D^{\alpha\alpha \beta \beta}_{\alpha\beta \alpha \beta}+{}^4D^{\alpha\alpha \beta \beta}_{\beta\alpha \alpha\beta}+{}^4D^{\alpha\alpha \beta \beta}_{\beta\alpha \beta\alpha}+{}^4D^{\alpha\alpha \beta \beta}_{\beta\beta \alpha\alpha}
\right. \nonumber\\
&\quad \quad \left.  + {}^4D^{\alpha\beta \beta \alpha}_{\alpha\alpha \beta \beta}+{}^4D^{\alpha\beta \beta \alpha}_{\alpha\beta \beta \alpha}+{}^4D^{\alpha\beta\beta \alpha}_{\alpha\beta \alpha \beta}+{}^4D^{\alpha\beta \beta \alpha}_{\beta\alpha \alpha\beta}+{}^4D^{\alpha\beta \beta \alpha}_{\beta\alpha \beta\alpha}+{}^4D^{\alpha\beta \beta \alpha}_{\beta\beta \alpha\alpha}\right.\nonumber\\
&\quad \quad \left.  + {}^4D^{\beta\alpha \beta \alpha}_{\alpha\alpha \beta \beta}+{}^4D^{\beta\alpha \beta \alpha}_{\alpha\beta \beta \alpha}+{}^4D^{\beta\alpha \beta \alpha}_{\alpha\beta \alpha \beta}+{}^4D^{\beta\alpha \beta \alpha}_{\beta\alpha \alpha\beta}+{}^4D^{\beta\alpha \beta \alpha}_{\beta\alpha \beta\alpha}+{}^4D^{\beta\alpha \beta \alpha}_{\beta\beta \alpha\alpha}\right.\nonumber\\
&\quad \quad \left.  + {}^4D^{\alpha \beta \alpha\beta}_{\alpha\alpha \beta \beta}+{}^4D^{\alpha \beta \alpha\beta}_{\alpha\beta \beta \alpha}+{}^4D^{\alpha \beta \alpha\beta}_{\alpha\beta \alpha \beta}+{}^4D^{\alpha \beta \alpha\beta}_{\beta\alpha \alpha\beta}+{}^4D^{\alpha \beta \alpha\beta}_{\beta\alpha \beta\alpha}+{}^4D^{\alpha \beta \alpha\beta}_{\beta\beta \alpha\alpha}\right.\nonumber\\
&\quad \quad \left.  + {}^4D^{\beta\alpha\alpha\beta}_{\alpha\alpha \beta \beta}+{}^4D^{\beta\alpha\alpha\beta}_{\alpha\beta \beta \alpha}+{}^4D^{\beta\alpha\alpha\beta}_{\alpha\beta \alpha \beta}+{}^4D^{\beta\alpha\alpha\beta}_{\beta\alpha \alpha\beta}+{}^4D^{\beta\alpha\alpha\beta}_{\beta\alpha \beta\alpha}+{}^4D^{\beta\alpha\alpha\beta}_{\beta\beta \alpha\alpha}\right.\nonumber\\
&\quad \quad \left.  + {}^4D^{\beta\beta\alpha\alpha}_{\alpha\alpha \beta \beta}+{}^4D^{\beta\beta\alpha\alpha}_{\alpha\beta \beta \alpha}+{}^4D^{\beta\beta\alpha\alpha}_{\alpha\beta \alpha \beta}+{}^4D^{\beta\beta\alpha\alpha}_{\beta\alpha \alpha\beta}+{}^4D^{\beta\beta\alpha\alpha}_{\beta\alpha \beta\alpha}+{}^4D^{\beta\beta\alpha\alpha}_{\beta\beta \alpha\alpha}
 \right) \nonumber\\
{}^4 D^{1 0}_{1 0} &=  \frac{3}{4}\left({}^4D^{\beta\alpha \beta \alpha}_{\beta\alpha \beta \alpha} - {}^4D^{\beta\alpha \beta \alpha}_{\alpha\beta \beta \alpha} +{}^4D^{\beta\alpha \beta \alpha}_{\beta\alpha \alpha \beta}-{}^4D^{\beta\alpha \beta \alpha}_{\alpha\beta \alpha\beta}
- {}^4D^{\alpha\beta \beta \alpha}_{\beta\alpha \beta \alpha} + {}^4D^{\alpha\beta \beta \alpha}_{\alpha\beta \beta \alpha} -{}^4D^{\alpha\beta \beta \alpha}_{\beta\alpha \alpha \beta}+{}^4D^{\alpha\beta \beta \alpha}_{\alpha\beta \alpha\beta}\right.\nonumber\\
&\quad\left.\quad + {}^4D^{\beta\alpha \alpha \beta}_{\beta\alpha \beta \alpha} - {}^4D^{\beta\alpha \alpha \beta}_{\alpha\beta \beta \alpha} +{}^4D^{\beta\alpha \alpha \beta}_{\beta\alpha \alpha \beta}-{}^4D^{\beta\alpha \alpha \beta}_{\alpha\beta \alpha\beta}
- {}^4D^{\alpha\beta \alpha\beta}_{\beta\alpha \beta \alpha} + {}^4D^{\alpha\beta \alpha\beta}_{\alpha\beta \beta \alpha} -{}^4D^{\alpha\beta \alpha\beta}_{\beta\alpha \alpha \beta}+{}^4D^{\alpha\beta \alpha\beta}_{\alpha\beta \alpha\beta} \right) \nonumber\\
{}^4 D^{0 0}_{0 0} &=\frac{1}{2}\left( {}^4D^{\beta\alpha\beta\alpha}_{\beta\alpha\beta\alpha} -{}^4D^{\beta\alpha\beta\alpha}_{\alpha\beta\beta\alpha}-{}^4D^{\beta\alpha\beta\alpha}_{\beta\alpha\alpha\beta}+{}^4D^{\beta\alpha\beta\alpha}_{\alpha\beta\alpha\beta}
-{}^4D^{\alpha\beta\beta\alpha}_{\beta\alpha\beta\alpha} +{}^4D^{\alpha\beta\beta\alpha}_{\alpha\beta\beta\alpha}+{}^4D^{\alpha\beta\beta\alpha}_{\beta\alpha\alpha\beta}-{}^4D^{\alpha\beta\beta\alpha}_{\alpha\beta\alpha\beta}
 \right.\nonumber\\
&\quad \left. \quad - {}^4D^{\beta\alpha\alpha\beta}_{\beta\alpha\beta\alpha} +{}^4D^{\beta\alpha\alpha\beta}_{\alpha\beta\beta\alpha}+{}^4D^{\beta\alpha\alpha\beta}_{\beta\alpha\alpha\beta}-{}^4D^{\beta\alpha\alpha\beta}_{\alpha\beta\alpha\beta}
+{}^4D^{\alpha\beta\alpha\beta}_{\beta\alpha\beta\alpha} -{}^4D^{\alpha\beta\alpha\beta}_{\alpha\beta\beta\alpha}-{}^4D^{\alpha\beta\alpha\beta}_{\beta\alpha\alpha\beta}+{}^4D^{\alpha\beta\alpha\beta}_{\alpha\beta\alpha\beta} \right)
  \nonumber\\
{}^4 D^{2 -1}_{2 -1} &=\frac{1}{4}\left({}^4D^{\beta\alpha\beta\beta}_{\beta\alpha\beta\beta} +{}^4D^{\beta\alpha\beta\beta}_{\alpha\beta\beta\beta} +{}^4D^{\beta\alpha\beta\beta}_{\beta\beta\beta\alpha}+{}^4D^{\beta\alpha\beta\beta}_{\beta\beta\alpha\beta}
+{}^4D^{\alpha\beta\beta\beta}_{\beta\alpha\beta\beta} +{}^4D^{\alpha\beta\beta\beta}_{\alpha\beta\beta\beta} +{}^4D^{\alpha\beta\beta\beta}_{\beta\beta\beta\alpha}+{}^4D^{\alpha\beta\beta\beta}_{\beta\beta\alpha\beta}  \right.\nonumber\\
&\quad\left. \quad +{}^4D^{\beta\beta\beta\alpha}_{\beta\alpha\beta\beta} +{}^4D^{\beta\beta\beta\alpha}_{\alpha\beta\beta\beta} +{}^4D^{\beta\beta\beta\alpha}_{\beta\beta\beta\alpha}+{}^4D^{\beta\beta\beta\alpha}_{\beta\beta\alpha\beta}
+ {}^4D^{\beta\beta\alpha\beta}_{\beta\alpha\beta\beta} +{}^4D^{\beta\beta\alpha\beta}_{\alpha\beta\beta\beta} +{}^4D^{\beta\beta\alpha\beta}_{\beta\beta\beta\alpha}+{}^4D^{\beta\beta\alpha\beta}_{\beta\beta\alpha\beta}
\right)\nonumber\\
{}^4 D^{1 -1}_{1 -1} &= \frac{3}{2}\left({}^4D^{\beta\alpha\beta\beta}_{\beta\alpha\beta\beta}-{}^4D^{\beta\alpha\beta\beta}_{\alpha\beta\beta\beta} -{}^4D^{\alpha\beta\beta\beta}_{\beta\alpha\beta\beta} +{}^4D^{\alpha\beta\beta\beta}_{\alpha\beta\beta\beta} \right)\nonumber\\
{}^4 D^{2 -2}_{2 -2} &= {}^4D^{\beta\beta\beta\beta}_{\beta\beta\beta\beta}\,.
\end{align}
Furthermore, the off-diagonal blocks, denoted as ${}^4D^{\lambda\mu}_{\lambda^\prime\mu}$ for $\lambda\neq \lambda^\prime$, are given by the products of the corresponding tensor operators ${}^4C^{\lambda\mu}$ and ${}^4C^{\lambda^\prime\mu}$.

\twocolumngrid
\bibliography{Refs}

\end{document}